\documentclass[sigconf]{acmart}
 
\usepackage{natbib}
\usepackage{multirow}
\usepackage{graphicx}
\usepackage{subcaption}
\usepackage{multicol}

\newcommand{\bumpup}{\vspace*{-.6ex}}

\newcommand{\items}{\mathcal{V}}

\setcopyright{acmlicensed}
\copyrightyear{2024}
\acmYear{2024}
\setcopyright{rightsretained}
\acmConference[RecSys '24]{18th ACM Conference on Recommender Systems}{October 14--18, 2024}{Bari, Italy}
\acmBooktitle{18th ACM Conference on Recommender Systems (RecSys '24), October 14--18, 2024, Bari, Italy}\acmDOI{10.1145/3640457.3691706}
\acmISBN{979-8-4007-0505-2/24/10}

\begin{document}

\title{Social Choice for Heterogeneous Fairness in Recommendation}

\author{Amanda Aird}
\email{amanda.aird@colorado.edu}
\affiliation{%
  \institution{Department of Information Science, University of Colorado, Boulder}
  \city{Boulder}
  \state{Colorado}
  \country{USA}
  \postcode{80309}
}

\author{Elena \u{S}tefancov\'{a}}
\email{elena.stefancova@fmph.uniba.sk}
\affiliation{%
  \institution{Comenius University Bratislava}
  \streetaddress{}
  \city{Bratislava}
  \country{Slovakia}
}

\author{Cassidy All}
\email{cassidy.all@colorado.edu}
\affiliation{%
  \institution{Department of Information Science, University of Colorado, Boulder}
  \city{Boulder}
  \state{Colorado}
  \country{USA}
  \postcode{80309}
}

\author{Amy Voida}
\email{amy.voida@colorado.edu}
\affiliation{%
  \institution{Department of Information Science, University of Colorado, Boulder}
  \city{Boulder}
  \state{Colorado}
  \country{USA}
  \postcode{80309}
}

\author{Martin Homola}
\email{homola@fmph.uniba.sk}
\affiliation{%
  \institution{Comenius University Bratislava}
  \streetaddress{}
  \city{Bratislava}
  \country{Slovakia}
}

\author{Nicholas Mattei}
\email{nsmattei@tulane.edu}
\affiliation{%
  \institution{Department of Computer Science, Tulane University}
  \city{New Orleans}
  \state{Louisiana}
  \country{USA}
  \postcode{70118}
}

\author{Robin Burke}
\email{robin.burke@colorado.edu}
\orcid{0000-0001-5766-6434}
\affiliation{%
  \institution{Department of Information Science, University of Colorado, Boulder}
  \city{Boulder}
  \state{Colorado}
  \country{USA}
  \postcode{80309}
}

\renewcommand{\shortauthors}{Aird, et al.}

\begin{abstract}
Algorithmic fairness in recommender systems requires close attention to the needs of a diverse set of stakeholders that may have competing interests. Previous work in this area has often been limited by fixed, single-objective definitions of fairness, built into algorithms or optimization criteria that are applied to a single fairness dimension or, at most, applied identically across dimensions. These narrow conceptualizations limit the ability to adapt fairness-aware solutions to the wide range of stakeholder needs and fairness definitions that arise in practice. Our work approaches recommendation fairness from the standpoint of computational social choice, using a multi-agent framework. In this paper, we explore the properties of different social choice mechanisms and demonstrate the successful integration of multiple, heterogeneous fairness definitions across multiple data sets. 
\end{abstract}

\ccsdesc[500]{Information systems~Recommender systems}
\ccsdesc[500]{Computing methodologies~Multi-agent systems}
\ccsdesc[500]{Social and professional topics~User characteristics}

\keywords{recommender systems, fairness, computational social choice}

\maketitle

\section{Introduction}

Fairness in algorithmic systems is a complex and multi-faceted area that is of research and practical concern. Different definitions may apply in different contexts, to different stakeholders, and in different types of applications. This complexity is particularly evident in recommender systems \cite{sonboli2022multisided} where there may be stakeholders of interest on the consumer or provider side of the recommendation interaction, or both \cite{burke_multisided_2017}. In the literature on fair recommender systems, a plethora of fairness definitions have emerged in the literature \cite{ekstrand2022fairness,smith2023scoping}, each with its own logic and associated algorithmic techniques. What nearly all of this literature has in common is two simplifying assumptions that are severely limiting to the potential applications of these ideas, and a barrier to deploying fair recommender systems in practice \cite{cramer2019challenges}.

The first limitation is an inability to capture the \textit{multiplicity} of fairness issues. With few exceptions \cite{wu2022multi,zehlike2022fair,aird2024dynamic}, fairness-aware recommender systems apply a single fairness definition to a single protected group, potentially one on each side of the interaction. This limitation is not realistic, as in most application contexts there will be multiple dimensions of fairness that need to be implemented. A second limitation is the assumption that a \textit{single fairness definition} will be appropriate, regardless of the protected group or the area of application. An architecture that assumes fairness will always take a certain form and measure is limited in its applicability \cite{selbst2019fairness}.

These limitations were addressed by the introduction of the Social Choice for Recommendation Fairness - Dynamic (SCRUF-D) architecture for fairness-aware re-ranking \cite{burke2022multi,aird2023exploring,aird2024dynamic}. In this paper, we extend prior results with two agents with the same underlying fairness definition to three agents with differing fairness definitions, showing that the system can support multiple heterogeneous fairness definitions simultaneously. We examine how different social choice mechanisms offer different trade-offs (and sometimes no trade-off) between fairness and accuracy.

\section{Related Work}
\label{sec:related}


Fairness in machine learning, especially in classification settings, is a popular topic, including formalizing definitions of fairness~\cite{chouldechova2017fair,dwork2012fairness,hardt2016equality,narayanan2018translation,Hutchinson_2019} and offering algorithmic techniques to mitigate unfairness \cite{kamiran2010discrimination,pedreshi2008discrimination,zemel2013learning,zhang2017anti}. However, despite these recent research efforts, the state-of-the-art offers little concrete guidance to industry practitioners  \cite{cramer2019challenges,holstein2019improving}. The unique problems of fairness in recommender systems have also been studied; see \citet{ekstrand2022fairness} for an overview. In recommendation, fairness concerns may arise on either the consumer side or on the provider side~\cite{burke_multisided_2017}. 
Other machine learning environments, such as classification, generally only need to consider the fairness properties of the system relative to the individuals being classified.


Both in the machine learning and the recommender systems formulation of fairness, there has been little recognition of the intersection of multiple fairness definitions and dimensions, although recent work has noted the benefits of combining multiple fairness definitions~\cite{beutel2019fairness,patro2022fair}. Most existing research considers only a single protected class. Even in cases where multiple groups are considered \cite{buolamwini2018gender,hebert2018multicalibration,kearns2018preventing,zhu2018fairness}, fairness is defined the same way for all groups. 


Many of the recent works on fair recommendation take a particular definition of fairness, and train a ranking model that includes fairness as a type of regularization on the subsequent loss function \cite{patro2020fairrec,suhr2019two,wu2022multi}. Such algorithms are brittle to changes in the definitions or measures of fairness. Post processing of the recommendations lists, often called re-ranking, is another popular method \cite{ferraro2021break} and one that we employ here for its ability to allow multiple and changeable definitions of fairness. 


\section{The SCRUF-D Architecture}\label{sec:formal}
The Social Choice for Recommendation Under Fairness - Dynamic (SCRUF-D) platform was introduced in \citet{aird2024dynamic}, as an extension to a simpler version found in \citet{sonboli2020and}. At a high level, SCRUF-D embodies the fairness concerns of multiple stakeholders as a set of \emph{fairness agents} that are \emph{allocated} to a user when recommendations are being generated. Once the agents are allocated to a user, their ordering of items is aggregated with the user's preferences (here coming from a traditional recommendation algorithm) by using a \emph{choice mechanism}, and the result is delivered to the user as their recommendations.

A fairness agent is defined by three functions: a fairness metric which is able to evaluate the history of the system and judge how fair the system has been to the agent over some time window, a compatibility metric which is the agent's evaluation of how much they want to be matched to a particular user, and a ranking function which expresses the agent's ordering/score of items. 

When a user arrives at the system, the set of agents express their preference for being matched with this user and their evaluation of how fair the system has been. These scores are passed to an \emph{allocation mechanism} which outputs a (randomized) allocation over the set of agents. Formally this is a probability distribution over the agents, and we examine different allocation functions. Once we have an allocation, the system generates an initial recommendation scoring of the items for the user using any standard recommendation algorithm. Each of the allocated fairness agents also express their ranking/scores of the items to give us a set of lists of items. These lists are then passed through a \emph{choice function}, which aggregates these lists into a final recommendation.

\section{Fairness Agents}

By necessity, our choice of fairness metrics is somewhat arbitrary. While our metrics are informed by research in particular application areas, we have not yet engaged in the full cycle of stakeholder consultation that would be required to formulate specific fairness metrics appropriate to each stakeholder group. Note also that we are examining only provider-side group fairness; consumer-side and individual fairness we leave to future work. 

The literature of provider-side fairness measures in recommender systems is extensive; see \citet{ekstrand2022fairness}. For the present study, we make use of fairness metrics that can be tied to a particular protected group. This excludes \textit{global fairness} measures that provide a score for the overall performance of the system relative to a fairness target distribution \footnote{An agent seeking fairness with respect to such measures could be implemented within SCRUF-D but it would be using a single definition of fairness on all groups, a different case from what we are considering here.}.

In this paper, we explore the following types of fairness metrics. Note that we are not proposing these classes of metrics as appropriate or desirable for any particular application of fairness-aware recommendation. The goal in formulating these metrics is to implement common but very different types of fairness metrics and demonstrate the ability of SCRUF-D to accommodate this heterogeneity across agents:

\textbf{Group Proportional Fairness ($m_{GPF}$):} Under group proportional fairness, we assume that there is a fixed proportion of recommendation results associated with the protected group that counts as a fair result from that group's perspective. We count the number of items that protected items appear across some set of recommendation lists, divide by the total number of recommendations and normalize by the desired target proportion, truncating at 1 once the target proportion is reached to maintain the 0..1 range.

\textbf{Group Utility Fairness ($m_{GUF}$):} While $m_{GPF}$ above captures the presence or absence of protected items in a list, it is indifferent to the item's position. However, highly-ranked positions may be of greater utility to providers. Also, $m_{GPF}$ does not take into account the \emph{number} of items that might be in each category. These considerations can be captured by modifying $m_{GPF}$ to sum rank-discounted utilities (here we discount by $log_2$ of the rank) rather than just count occurrences, and to normalize the utility for protected and unprotected groups by their respective sizes. 

\textbf{Group MRR Fairness ($m_{MRR}$):} The measures above look at all recommended items in a set of lists summing a total value for protected items. In some cases it might be desirable to focus on a minimal degree of representation across recommendation lists, and we capture this type of metric using \textit{mean reciprocal rank}. We average the reciprocal rank of the highest ranking protected item across all lists and normalize by the target MRR value. 

\section{Methodology}\label{sec:methods}

We make use of two data sets MovieLens and Microlending. For each data set, we define three sensitive attributes and protected values and define an agent for each using metrics based on the three different definitions above. The MovieLens 1M dataset \cite{movielens} contains user ratings for movies with has 3,900 movies, 6,040 users, and approximately 1 million ratings. We consider the sensitive attributes in this dataset to be (1) movies with at least one female writer and/or director, (2) movies with non-English scripts, and (3) older movies (released before 1990). Our second dataset is the Microlending 2017 dataset \cite{sonboli2022micro}. This dataset includes 2,673 pseudo-items\footnote{The Microlending 2017 dataset represents user preferences relative to clusters of items (pseudo-items) in place of individual loans since loans in this dataset have small numbers of associated users and the unclustered data is too sparse for collaborative recommendation. See \cite{sonboli2022micro} for additional details.}, 4,005 lenders and 110,371 ratings. We consider the sensitive attributes in this dataset to be (1) loans from countries with low funding rates, (2) loans funding sectors that have low funding rates, and (3) loans larger than \$5,000. 

\subsection{Agent Definitions}
As noted in Section \ref{sec:formal}, fairness agents in SCRUF-D are defined by a fairness metric, a compatibility metric, and a ranking function. In this study, we concentrate on the fairness metric. 

We define the compatibility between a user and an item feature as the frequency of occurrence of that feature among the items that the user likes. For the Microlending dataset, all funded loans are considered liked; in the Movies dataset, we categorize any movie with a rating over 3 as liked. Compatibility is calculated as $c_{u, f} = p_{u, f}/\bar{p}_f$ where 
$p_{u, i} = \text{{count of liked items with feature f}}/\text{{total count of items rated}}$
and $\bar{p}_f$ is the average number of items with feature $f$. The compatibility scores are normalized across features for each user. 

For ranking, we use two approaches depending on the type of choice mechanism. For the \emph{Rescoring Mechanism}, we use a simple binary partial order $\items_s > \items_{\neg s}$ where the agent prefers all items that it considers as protected. For our other choice mechanisms, this two-level partial ranking works poorly as it induces many ties, forcing the system to rely on an arbitrary tie-breaking rule. To address this issue, we implement a \textit{cascaded} ranking function that integrates both the protected class preference and inherits, as a secondary ranking criterion, the ranking of the recommender system agent. This is effectively the same as moving all of the protected items to the front of the recommendation list, keeping the between-item preferences the same. This method removes ties, inducing a total order. 

For each of our datasets we define three different fairness agents, with three different metrics, to explore the flexibility and trade-offs of SCRUF-D. For the Microlending experiments, the first agent is focused on loans in amounts greater than \$5,000, which are considered to have proportionally stronger economic impact but also tend to be funded at a lower rate. This agent aims to have such loans represented across the recommendation lists with MRR target value of 0.5. The second agent is focused on loans from countries with low funding rates. The aim for this agent is to ensure higher utility for loans for countries with historically low funding rates. The third agent focuses on loans from sectors with low funding rates. This agent uses the proportional definition of fairness; the goal is that 20\% of all loans recommended are from sectors with historically low funding rates.
For the Movie data experiments, the first agent is focused on movies with women writers and directors. This agent is to ensure that a movie with a woman writer and/or director is in the top 1--2 positions of users' lists. The second agent is focused on non-English movies. This agent has a goal of ensuring non-English movies receive equal utility compared to movies in English. The third agent is focused on older movies, with the goal of making 25\% of all recommended movies older movies. 

\begin{figure*}[th]
   \centering
      \begin{subfigure}{0.45\textwidth}
       \centering
       \includegraphics[width=0.8\linewidth]{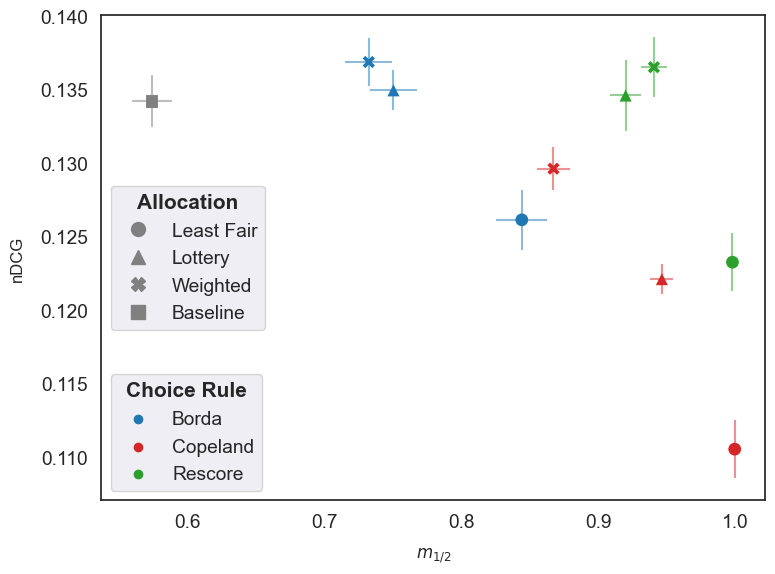}
       \Description[A scatterplot with confidence intervals on X and Y]{A scatterplot with points of each combination of allocation and choice mechanism. The dots (other than baseline) occupy the upper right half of the image.}
       \caption{Microlending data}
        \label{fig:kiva-scatter}
   \end{subfigure}
   \begin{subfigure}{0.45\textwidth}
       \centering
       \includegraphics[width=0.8\linewidth]{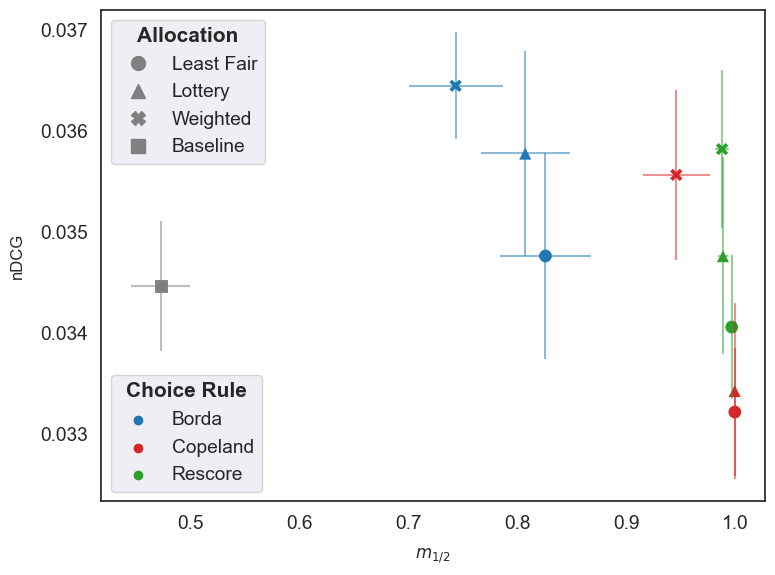}
       \Description[A scatterplot with confidence intervals on X and Y]{A scatterplot with points of each combination of allocation and choice mechanism. The dots (other than baseline) occupy the right half of the image and the confidence intervals are rel\Description[A scatterplot with confidence intervals on X and Y]{A scatterplot with points of each combination of allocation and choice mechanism. The dots (other than baseline) occupy the upper right half of the image.}atively large.}
       \caption{Movies data}
          \label{fig:movie-scatter}
   \end{subfigure}
   \caption{nDCG vs $l_{1/2}$ Fairness Norm for Microlending (Left) and Movies (Right).}
   \label{fig:scatter}
   \bumpup
\end{figure*}

\subsection{Mechanisms}
Instantiating the SCRUF-D system requires choosing mechanisms that will be applied in the allocation and choice phases. There are a wide variety of options for each of these tasks. For the purposes of this study, we concentrate on three allocation mechanisms:

\textbf{Least Fair}: This mechanism allocates a single agent for each recommendation opportunity by comparing the fairness metric $m_i$ for each agent and allocating the agent with the lowest value. If all agents have a fairness of 1.0 (the target), then no agents are allocated to the arriving user(s). This mechanism ignores the compatibility computation, which optimizes fairness but as we show, ignoring the user's preferences results in a greater loss of accuracy.

\textbf{Lottery}: This mechanism allocates only a single agent  using a non-deterministic allocation \cite{brandt2017rolling} over the set of agents using a probability distribution. The distribution is computed by calculating the product of unfairness (1 - fairness) and compatibility, where each is raised to a small integer power. An agent is to a recommendation opportunity with high probability if fairness need and the compatibility are high, and the exponentiation allows the product to be tuned. In our experiments we set the power of compatibility to $2$, which discounts it somewhat compared to fairness. These scores are computed over all agents and normalized to sum to 1. 

\textbf{Weighted}: The weighted mechanism uses the same distribution as calculated for the Lottery mechanism, but instead of selecting only a single agent, all agents with fairnness $ < 1.0$ are allocated and weighted according to their value in the distribution. 

These allocation mechanisms allow us to explore two key aspects of the SCRUF-D platform. First, by comparing Least Fair against the other mechanisms, we can see the value of considering user characteristics in agent allocation. Second, by comparing the Lottery vs Weighted schemes, we can see the difference between sending all of the agents to the choice phase as opposed to allocating a single agent. Since the scoring and re-ranking processes can be computationally intensive, there is an advantage to having only two agents in the choice phase if there is no cost in performance.

In the choice phase, we have again a wide variety of preference aggregation schemes to choose from. SCRUF-D integrates Whalrus\footnote{https://github.com/francois-durand/whalrus}, a well-known library implementing a variety of voting rules. From the available methods, this paper explores three:

\textbf{Borda}: The Borda method \cite{Zwicker:Voting} assigns a score to each rank and sums these scores for every item to achieve a final scoring / ranking. To allow our implementation to be tuned for the best trade-off between fairness and accuracy, we use a weighted version of the method with the recommender weight set to 0.6. This gives the recommender 1.5x as much weight in the final outcome as compared to the allocated agents. 

\textbf{Copeland}: The Copeland mechanism is a pair-wise method. It uses the win-loss record for each pair of items on each the ballot, and awards each item a point per win. These scores are then used to order the items \cite{sep-voting-methods}. We use a weighted version; as with Borda, this is realized by multiplying the number of ballots. 

\textbf{Rescoring}: The Rescoring mechanism makes use of the scores from the recommendation function, rather than only rankings. A linear combination of the scores from each allocated agent (and the recommender) are computed and the items are reordered based on these values. In our scoring function, an agent contributes a fixed $\delta = 0.5$ to the score for protected items and 0 for unprotected items\footnote{This $\delta$ value was found to provide a good fairness / accuracy trade-off in prior work.}. The values are weighted first by the agent's allocation weight and then by the inverse of the weight associated with the recommender. 

These mechanisms allow us to explore several dimensions of preference aggregation. First, there is the pair-wise versus scoring-based distinction with Copeland, the pair-wise option. Second, there is the difference between ordinal social choice methods and the Rescoring method that makes use of the real-valued predicted ratings arising from the recommendation function.

Layering the allocation and choice mechanisms onto an existing recommender system introduces a small computational overhead (complexity), however, we see the risks of integrating a single fairness definition into the recommender (which would need to be completely retrained for a new metric) as far outweighing the (computational) cost. In terms of the allocation stage, each of the proposed mechanisms can be computed from the lottery history in $\mathcal{O}(n)$ time where $n$ is the length of the history to be considered. This computation is bounded by the window size and can scale depending on how reactive we want the system to be. For the choice mechanism, Borda and Rescoring can be computed in $\mathcal{O}(n)$ while Copeland is $\mathcal{O}(n^2)$ where $n$ is the number of elements in the output list \cite{Zwicker:Voting}. We expect in most practical applications for both of these to be smaller numbers which do not incur significant overhead.

\subsection{Evaluation}
We evaluate the results of these experiments examining fairness and accuracy. All results were computed over five training / test folds of the data and averaged. Fairness is computed relative to the metric associated with each agent. However, this summative fairness score is over the entire experiment, not the  time-bounded window that agents compute over during experiment execution. 

We summarize the fairness scores across all the agents using a function of the $l_{1/2}$ norm of over the fairness scores for each individual agent \cite{mehrotra2018towards}. This metric is maximized (and equal to the mean) when all of the scores are equal, and it is below the mean when the scores are unequal. The value is normalized to $[0,1]$.

Accuracy is computed using nDCG based on held-out test data in each data set. For the purposes of this study, we did not use any other algorithms as comparators; we are only comparing SCRUF-D's output against the non-re-ranked results. As noted in Section~\ref{sec:related}, there are very few algorithms that attempt to address multiple fairness concerns simultaneously and only one, OFair \cite{sonboli2020opportunistic}, is capable of supporting heterogeneous fairness definitions. Prior work found that OFair is not competitive against SCRUF-D \cite{aird2024dynamic}.  

\section{Results}

Figure~\ref{fig:kiva-scatter} shows the results for the Microlending data comparing accuracy (x-axis) with the $l_{1/2}$ fairness norm on the y-axis. (Individual fairness results for each agent are not included for reasons of space.) Although Least Fair has very strong fairness outcomes across different options for the choice mechanism, it suffers from low accuracy. This matches our expectation that taking user compatibility into account allows the system to make a better trade-off. 

For the Borda and Rescore choice mechanisms, the Lottery and Weighted allocations look quite similar considering their 95\% confidence intervals, suggesting that these allocation mechanisms can give similar results. This is encouraging because the Lottery has the potential for greater efficiency.

Across the choice mechanisms, Borda tends to have the smallest fairness improvement and does a somewhat better job of preserving accuracy. Copeland offers improved fairness, but these choice mechanisms are dominated by  Rescore, which achieves better than 0.9 on the $l_{1/2}$ metric without a noticeable loss in accuracy. 

Figure~\ref{fig:movie-scatter} shows the positioning of the different mechanism combinations in the accuracy / fairness space for the Movies dataset. We still see Rescoring as dominant and Borda in a lower fairness/higher accuracy position, with Copeland somewhere in the middle. We also see the Rescore and Copeland mechanisms clustered at the far right indicating that they are able to reach the fairness targets across all the agents almost fully.

One interesting difference is the positioning of the Lottery and Weighted mechanisms. In the Microlending data, these mechanisms gave similar results with Lottery a bit lower in accuracy. In the Movies results, the difference is more pronounced and there is greater cost for the Lottery mechanism. On the other hand, the accuracy values are closer to baseline in Movies, showing that high fairness does not have to come at the cost of accuracy loss.

\section{Conclusion}
This paper addresses two key limitations in existing fairness-aware recommendation research that have not been addressed in prior work. First, we formulate multiple definitions of provider-side fairness relevant to real-world datasets. Prior work has been limited to, at most two concerns, usually on different sides of the recommendation interaction. Second, the work posits heterogeneous fairness definitions, allowing different fairness issues to be represented by different metrics. We believe that allowing a multiplicity of concerns and allowing for varied fairness definitions and targets is essential in practical settings.  

In the context of computational social choice, we show that SCRUF-D is compatible with a range of different allocation and choice mechanisms and, while some general patterns can be seen, that these mechanisms work differently across recommendation domains and datasets. We find that in some datasets a Lottery mechanism can be competitive with one that allocates multiple fairness agents at a time, suggesting potential efficiency in applying these techniques in practice.

Numerous additional challenges remain. This work has concentrated on provider-side group fairness under multiple definitions. Additional definitions exist and are worth exploring. There is also the question of scale: how many simultaneous agents can be supported? In addition, we believe that SCRUF-D is capable of supporting individual fairness and consumer-side fairness, but additional work needs to be done to demonstrate this capacity. 

\begin{acks}
Burke, Voida and Aird were supported by the National Science Foundation under grant awards IIS-1911025 and IIS-2107577. Mattei was supported by NSF Grant IIS-2107505. Stefancova was supported by Slovak Research and Development Agency under Contract no. APVV-20-0353 and the Fulbright program.
\end{acks}

\bibliographystyle{ACM-Reference-Format}
\bibliography{scruf}


\begin{thebibliography}{38}


\ifx \showCODEN    \undefined \def \showCODEN     #1{\unskip}     \fi
\ifx \showDOI      \undefined \def \showDOI       #1{#1}\fi
\ifx \showISBNx    \undefined \def \showISBNx     #1{\unskip}     \fi
\ifx \showISBNxiii \undefined \def \showISBNxiii  #1{\unskip}     \fi
\ifx \showISSN     \undefined \def \showISSN      #1{\unskip}     \fi
\ifx \showLCCN     \undefined \def \showLCCN      #1{\unskip}     \fi
\ifx \shownote     \undefined \def \shownote      #1{#1}          \fi
\ifx \showarticletitle \undefined \def \showarticletitle #1{#1}   \fi
\ifx \showURL      \undefined \def \showURL       {\relax}        \fi
\providecommand\bibfield[2]{#2}
\providecommand\bibinfo[2]{#2}
\providecommand\natexlab[1]{#1}
\providecommand\showeprint[2][]{arXiv:#2}

\bibitem[Aird et~al\mbox{.}(2023)]%
        {aird2023exploring}
\bibfield{author}{\bibinfo{person}{Amanda Aird}, \bibinfo{person}{Cassidy All}, \bibinfo{person}{Paresha Farastu}, \bibinfo{person}{Elena Stefancova}, \bibinfo{person}{Joshua Sun}, \bibinfo{person}{Nicholas Mattei}, {and} \bibinfo{person}{Robin Burke}.} \bibinfo{year}{2023}\natexlab{}.
\newblock \bibinfo{title}{Exploring Social Choice Mechanisms for Recommendation Fairness in SCRUF}.
\newblock \bibinfo{howpublished}{Presented at the 2023 FAccTRec Workshop on Responsible Recommendation.}.
\newblock
\showeprint[arxiv]{2309.08621}~[cs.IR]
\urldef\tempurl%
\url{https://arxiv.org/abs/2309.08621}
\showURL{%
\tempurl}


\bibitem[Aird et~al\mbox{.}(2024)]%
        {aird2024dynamic}
\bibfield{author}{\bibinfo{person}{Amanda Aird}, \bibinfo{person}{Paresha Farastu}, \bibinfo{person}{Joshua Sun}, \bibinfo{person}{Elena Štefancová}, \bibinfo{person}{Cassidy All}, \bibinfo{person}{Amy Voida}, \bibinfo{person}{Nicholas Mattei}, {and} \bibinfo{person}{Robin Burke}.} \bibinfo{year}{2024}\natexlab{}.
\newblock \bibinfo{title}{Dynamic fairness-aware recommendation through multi-agent social choice}.
\newblock
\newblock
\showeprint[arxiv]{2303.00968}~[cs.AI]
\urldef\tempurl%
\url{https://arxiv.org/abs/2303.00968}
\showURL{%
\tempurl}


\bibitem[Beutel et~al\mbox{.}(2019)]%
        {beutel2019fairness}
\bibfield{author}{\bibinfo{person}{Alex Beutel}, \bibinfo{person}{Jilin Chen}, \bibinfo{person}{Tulsee Doshi}, \bibinfo{person}{Hai Qian}, \bibinfo{person}{Li Wei}, \bibinfo{person}{Yi Wu}, \bibinfo{person}{Lukasz Heldt}, \bibinfo{person}{Zhe Zhao}, \bibinfo{person}{Lichan Hong}, \bibinfo{person}{Ed~H. Chi}, {and} \bibinfo{person}{Cristos Goodrow}.} \bibinfo{year}{2019}\natexlab{}.
\newblock \bibinfo{title}{Fairness in Recommendation Ranking through Pairwise Comparisons}.
\newblock , \bibinfo{numpages}{16}~pages.
\newblock
\showeprint[arxiv]{1903.00780}~[cs.CY]
\urldef\tempurl%
\url{https://arxiv.org/abs/1903.00780}
\showURL{%
\tempurl}


\bibitem[Brandt(2017)]%
        {brandt2017rolling}
\bibfield{author}{\bibinfo{person}{Felix Brandt}.} \bibinfo{year}{2017}\natexlab{}.
\newblock \showarticletitle{Rolling the dice: Recent results in probabilistic social choice}.
\newblock In \bibinfo{booktitle}{\emph{Trends in Computational Social Choice}}, \bibfield{editor}{\bibinfo{person}{Ulle Endriss}} (Ed.). \bibinfo{publisher}{AI Access}, Chapter~1, \bibinfo{pages}{3--26}.
\newblock


\bibitem[Buolamwini and Gebru(2018)]%
        {buolamwini2018gender}
\bibfield{author}{\bibinfo{person}{Joy Buolamwini} {and} \bibinfo{person}{Timnit Gebru}.} \bibinfo{year}{2018}\natexlab{}.
\newblock \showarticletitle{Gender shades: Intersectional accuracy disparities in commercial gender classification}. In \bibinfo{booktitle}{\emph{Conference on Fairness, Accountability and Transparency}}. \bibinfo{publisher}{ACM}, \bibinfo{pages}{77--91}.
\newblock


\bibitem[Burke(2017)]%
        {burke_multisided_2017}
\bibfield{author}{\bibinfo{person}{Robin Burke}.} \bibinfo{year}{2017}\natexlab{}.
\newblock \showarticletitle{Multisided {Fairness} for {Recommendation}}. In \bibinfo{booktitle}{\emph{Workshop on {Fairness}, {Accountability} and {Transparency} in {Machine} {Learning} ({FATML})}}. \bibinfo{address}{Halifax, Nova Scotia}, \bibinfo{numpages}{5}~pages.
\newblock
\urldef\tempurl%
\url{https://arxiv.org/abs/1707.00093}
\showURL{%
\tempurl}


\bibitem[Burke et~al\mbox{.}(2022)]%
        {burke2022multi}
\bibfield{author}{\bibinfo{person}{Robin Burke}, \bibinfo{person}{Nicholas Mattei}, \bibinfo{person}{Vladislav Grozin}, \bibinfo{person}{Amy Voida}, {and} \bibinfo{person}{Nasim Sonboli}.} \bibinfo{year}{2022}\natexlab{}.
\newblock \showarticletitle{Multi-agent Social Choice for Dynamic Fairness-aware Recommendation}. In \bibinfo{booktitle}{\emph{Adjunct Proceedings of the 30th ACM Conference on User Modeling, Adaptation and Personalization}}. \bibinfo{publisher}{ACM}, \bibinfo{pages}{234--244}.
\newblock


\bibitem[Chouldechova(2017)]%
        {chouldechova2017fair}
\bibfield{author}{\bibinfo{person}{Alexandra Chouldechova}.} \bibinfo{year}{2017}\natexlab{}.
\newblock \showarticletitle{Fair prediction with disparate impact: A study of bias in recidivism prediction instruments}.
\newblock \bibinfo{journal}{\emph{Big data}} \bibinfo{volume}{5}, \bibinfo{number}{2} (\bibinfo{year}{2017}), \bibinfo{pages}{153--163}.
\newblock


\bibitem[Cramer et~al\mbox{.}(2019)]%
        {cramer2019challenges}
\bibfield{author}{\bibinfo{person}{Henriette Cramer}, \bibinfo{person}{Kenneth Holstein}, \bibinfo{person}{Jennifer~W. Vaughan}, \bibinfo{person}{Hal Daum\'{e}~III}, \bibinfo{person}{Miroslav Dud\'{i}k}, \bibinfo{person}{Hanna Wallach}, \bibinfo{person}{Sravana Reddy}, {and} \bibinfo{person}{Jean Garcia-Gathright}.} \bibinfo{year}{2019}\natexlab{}.
\newblock \bibinfo{title}{Challenges of incorporating algorithmic fairness into industry practice}.
\newblock
\newblock


\bibitem[Dwork et~al\mbox{.}(2012)]%
        {dwork2012fairness}
\bibfield{author}{\bibinfo{person}{Cynthia Dwork}, \bibinfo{person}{Moritz Hardt}, \bibinfo{person}{Toniann Pitassi}, \bibinfo{person}{Omer Reingold}, {and} \bibinfo{person}{Richard Zemel}.} \bibinfo{year}{2012}\natexlab{}.
\newblock \showarticletitle{Fairness through awareness}. In \bibinfo{booktitle}{\emph{Proceedings of the 3rd Innovations in Theoretical Computer Science Conference}}. ACM, \bibinfo{pages}{214--226}.
\newblock


\bibitem[Ekstrand et~al\mbox{.}(2022)]%
        {ekstrand2022fairness}
\bibfield{author}{\bibinfo{person}{Michael~D. Ekstrand}, \bibinfo{person}{Anubrata Das}, \bibinfo{person}{Robin Burke}, {and} \bibinfo{person}{Fernando Diaz}.} \bibinfo{year}{2022}\natexlab{}.
\newblock \bibinfo{title}{Fairness in Information Access Systems}.
\newblock
\newblock
\showeprint[arxiv]{2105.05779}~[cs.IR]


\bibitem[Ferraro et~al\mbox{.}(2021)]%
        {ferraro2021break}
\bibfield{author}{\bibinfo{person}{Andres Ferraro}, \bibinfo{person}{Xavier Serra}, {and} \bibinfo{person}{Christine Bauer}.} \bibinfo{year}{2021}\natexlab{}.
\newblock \showarticletitle{Break the loop: Gender imbalance in music recommenders}. In \bibinfo{booktitle}{\emph{Proceedings of the 2021 Conference on Human Information Interaction and Retrieval}}. \bibinfo{pages}{249--254}.
\newblock


\bibitem[Hardt et~al\mbox{.}(2016)]%
        {hardt2016equality}
\bibfield{author}{\bibinfo{person}{Moritz Hardt}, \bibinfo{person}{Eric Price}, {and} \bibinfo{person}{Nati Srebro}.} \bibinfo{year}{2016}\natexlab{}.
\newblock \showarticletitle{Equality of opportunity in supervised learning}. In \bibinfo{booktitle}{\emph{Advances in neural information processing systems}}. \bibinfo{pages}{3315--3323}.
\newblock


\bibitem[Harper and Konstan(2015)]%
        {movielens}
\bibfield{author}{\bibinfo{person}{F~Maxwell Harper} {and} \bibinfo{person}{Joseph~A Konstan}.} \bibinfo{year}{2015}\natexlab{}.
\newblock \showarticletitle{The MovieLens Datasets: History and Context}.
\newblock \bibinfo{journal}{\emph{ACM Transactions on Interactive Intelligent Systems (TiiS)}} \bibinfo{volume}{5}, \bibinfo{number}{4} (\bibinfo{year}{2015}), \bibinfo{pages}{19}.
\newblock


\bibitem[H{\'e}bert-Johnson et~al\mbox{.}(2018)]%
        {hebert2018multicalibration}
\bibfield{author}{\bibinfo{person}{{\'U}rsula H{\'e}bert-Johnson}, \bibinfo{person}{Michael Kim}, \bibinfo{person}{Omer Reingold}, {and} \bibinfo{person}{Guy Rothblum}.} \bibinfo{year}{2018}\natexlab{}.
\newblock \showarticletitle{Multicalibration: Calibration for the (Computationally-identifiable) masses}. In \bibinfo{booktitle}{\emph{International Conference on Machine Learning}}. \bibinfo{pages}{1944--1953}.
\newblock


\bibitem[Holstein et~al\mbox{.}(2019)]%
        {holstein2019improving}
\bibfield{author}{\bibinfo{person}{Kenneth Holstein}, \bibinfo{person}{Jennifer~W. Vaughan}, \bibinfo{person}{Hal Daum\'{e}~III}, \bibinfo{person}{Miroslav Dud\'{i}k}, {and} \bibinfo{person}{Hanna Wallach}.} \bibinfo{year}{2019}\natexlab{}.
\newblock \showarticletitle{Improving fairness in machine learning systems: What do industry practitioners need?}. In \bibinfo{booktitle}{\emph{Proceedings of the Conference of Human Computer Interaction (CHI'19)}}. ACM, \bibinfo{numpages}{16}~pages.
\newblock
\urldef\tempurl%
\url{arXiv:1812.05239v2 [cs.HC] 7 Jan 2019}
\showURL{%
\tempurl}


\bibitem[Hutchinson and Mitchell(2019)]%
        {Hutchinson_2019}
\bibfield{author}{\bibinfo{person}{Ben Hutchinson} {and} \bibinfo{person}{Margaret Mitchell}.} \bibinfo{year}{2019}\natexlab{}.
\newblock \showarticletitle{50 years of test (un) fairness: Lessons for machine learning}. In \bibinfo{booktitle}{\emph{Proceedings of the conference on fairness, accountability, and transparency}}. \bibinfo{publisher}{ACM}, \bibinfo{pages}{49--58}.
\newblock


\bibitem[Kamiran et~al\mbox{.}(2010)]%
        {kamiran2010discrimination}
\bibfield{author}{\bibinfo{person}{Faisal Kamiran}, \bibinfo{person}{Toon Calders}, {and} \bibinfo{person}{Mykola Pechenizkiy}.} \bibinfo{year}{2010}\natexlab{}.
\newblock \showarticletitle{Discrimination aware decision tree learning}. In \bibinfo{booktitle}{\emph{Data Mining (ICDM), 2010 IEEE 10th International Conference on}}. IEEE, \bibinfo{pages}{869--874}.
\newblock


\bibitem[Kearns et~al\mbox{.}(2018)]%
        {kearns2018preventing}
\bibfield{author}{\bibinfo{person}{Michael Kearns}, \bibinfo{person}{Seth Neel}, \bibinfo{person}{Aaron Roth}, {and} \bibinfo{person}{Zhiwei~Steven Wu}.} \bibinfo{year}{2018}\natexlab{}.
\newblock \bibinfo{title}{Preventing Fairness Gerrymandering: Auditing and Learning for Subgroup Fairness}.
\newblock
\newblock
\showeprint[arxiv]{1711.05144}~[cs.LG]


\bibitem[Mehrotra et~al\mbox{.}(2018)]%
        {mehrotra2018towards}
\bibfield{author}{\bibinfo{person}{Rishabh Mehrotra}, \bibinfo{person}{James McInerney}, \bibinfo{person}{Hugues Bouchard}, \bibinfo{person}{Mounia Lalmas}, {and} \bibinfo{person}{Fernando Diaz}.} \bibinfo{year}{2018}\natexlab{}.
\newblock \showarticletitle{Towards a Fair Marketplace: Counterfactual Evaluation of the Trade-off between Relevance, Fairness \& Satisfaction in Recommendation Systems}. In \bibinfo{booktitle}{\emph{Proceedings of the Conference on Information and Knowledge Management}}. \bibinfo{pages}{2243--2251}.
\newblock


\bibitem[Narayanan(2018)]%
        {narayanan2018translation}
\bibfield{author}{\bibinfo{person}{Arvind Narayanan}.} \bibinfo{year}{2018}\natexlab{}.
\newblock \showarticletitle{Translation tutorial: 21 fairness definitions and their politics}. In \bibinfo{booktitle}{\emph{Proc. Conf. Fairness Accountability Transp., New York, USA}}, Vol.~\bibinfo{volume}{1170}. \bibinfo{publisher}{ACM}, \bibinfo{pages}{3}.
\newblock


\bibitem[Pacuit(2019)]%
        {sep-voting-methods}
\bibfield{author}{\bibinfo{person}{Eric Pacuit}.} \bibinfo{year}{2019}\natexlab{}.
\newblock \showarticletitle{{Voting Methods}}.
\newblock In \bibinfo{booktitle}{\emph{The {Stanford} Encyclopedia of Philosophy} (\bibinfo{edition}{{F}all 2019} ed.)}, \bibfield{editor}{\bibinfo{person}{Edward~N. Zalta}} (Ed.). \bibinfo{publisher}{Metaphysics Research Lab, Stanford University}.
\newblock


\bibitem[Patro et~al\mbox{.}(2020)]%
        {patro2020fairrec}
\bibfield{author}{\bibinfo{person}{Gourab~K Patro}, \bibinfo{person}{Arpita Biswas}, \bibinfo{person}{Niloy Ganguly}, \bibinfo{person}{Krishna~P Gummadi}, {and} \bibinfo{person}{Abhijnan Chakraborty}.} \bibinfo{year}{2020}\natexlab{}.
\newblock \showarticletitle{FairRec: Two-Sided Fairness for Personalized Recommendations in Two-Sided Platforms}. In \bibinfo{booktitle}{\emph{Proceedings of The Web Conference 2020}}. \bibinfo{pages}{1194--1204}.
\newblock


\bibitem[Patro et~al\mbox{.}(2022)]%
        {patro2022fair}
\bibfield{author}{\bibinfo{person}{Gourab~K Patro}, \bibinfo{person}{Lorenzo Porcaro}, \bibinfo{person}{Laura Mitchell}, \bibinfo{person}{Qiuyue Zhang}, \bibinfo{person}{Meike Zehlike}, {and} \bibinfo{person}{Nikhil Garg}.} \bibinfo{year}{2022}\natexlab{}.
\newblock \showarticletitle{Fair ranking: a critical review, challenges, and future directions}. In \bibinfo{booktitle}{\emph{Proceedings of the 2022 ACM Conference on Fairness, Accountability, and Transparency}}. \bibinfo{pages}{1929--1942}.
\newblock


\bibitem[Pedreshi et~al\mbox{.}(2008)]%
        {pedreshi2008discrimination}
\bibfield{author}{\bibinfo{person}{Dino Pedreshi}, \bibinfo{person}{Salvatore Ruggieri}, {and} \bibinfo{person}{Franco Turini}.} \bibinfo{year}{2008}\natexlab{}.
\newblock \showarticletitle{Discrimination-aware data mining}. In \bibinfo{booktitle}{\emph{Proceedings of the 14th ACM SIGKDD international conference on Knowledge discovery and data mining}}. ACM, \bibinfo{pages}{560--568}.
\newblock


\bibitem[Selbst et~al\mbox{.}(2019)]%
        {selbst2019fairness}
\bibfield{author}{\bibinfo{person}{Andrew~D Selbst}, \bibinfo{person}{Danah Boyd}, \bibinfo{person}{Sorelle~A Friedler}, \bibinfo{person}{Suresh Venkatasubramanian}, {and} \bibinfo{person}{Janet Vertesi}.} \bibinfo{year}{2019}\natexlab{}.
\newblock \showarticletitle{Fairness and abstraction in sociotechnical systems}. In \bibinfo{booktitle}{\emph{Proceedings of the conference on fairness, accountability, and transparency}}. \bibinfo{publisher}{ACM}, \bibinfo{address}{New York}, \bibinfo{pages}{59--68}.
\newblock


\bibitem[Smith et~al\mbox{.}(2023)]%
        {smith2023scoping}
\bibfield{author}{\bibinfo{person}{Jessie~J Smith}, \bibinfo{person}{Lex Beattie}, {and} \bibinfo{person}{Henriette Cramer}.} \bibinfo{year}{2023}\natexlab{}.
\newblock \showarticletitle{Scoping Fairness Objectives and Identifying Fairness Metrics for Recommender Systems: The Practitioners’ Perspective}. In \bibinfo{booktitle}{\emph{Proceedings of the ACM Web Conference 2023}}. \bibinfo{pages}{3648--3659}.
\newblock


\bibitem[Sonboli et~al\mbox{.}(2022a)]%
        {sonboli2022micro}
\bibfield{author}{\bibinfo{person}{Nasim Sonboli}, \bibinfo{person}{Amanda Aird}, {and} \bibinfo{person}{Robin Burke}.} \bibinfo{year}{2022}\natexlab{a}.
\newblock \bibinfo{booktitle}{\emph{Microlending 2017 Data Set}}.
\newblock University of Colorado Boulder.
\newblock
\urldef\tempurl%
\url{https://doi.org/10.25810/PGJK-RR19}
\showDOI{\tempurl}


\bibitem[Sonboli et~al\mbox{.}(2022b)]%
        {sonboli2022multisided}
\bibfield{author}{\bibinfo{person}{Nasim Sonboli}, \bibinfo{person}{Robin Burke}, \bibinfo{person}{Michael Ekstrand}, {and} \bibinfo{person}{Rishabh Mehrotra}.} \bibinfo{year}{2022}\natexlab{b}.
\newblock \showarticletitle{The multisided complexity of fairness in recommender systems}.
\newblock \bibinfo{journal}{\emph{AI magazine}} \bibinfo{volume}{43}, \bibinfo{number}{2} (\bibinfo{year}{2022}), \bibinfo{pages}{164--176}.
\newblock


\bibitem[Sonboli et~al\mbox{.}(2020a)]%
        {sonboli2020and}
\bibfield{author}{\bibinfo{person}{Nasim Sonboli}, \bibinfo{person}{Robin Burke}, \bibinfo{person}{Nicholas Mattei}, \bibinfo{person}{Farzad Eskandanian}, {and} \bibinfo{person}{Tian Gao}.} \bibinfo{year}{2020}\natexlab{a}.
\newblock \bibinfo{title}{"And the Winner Is...": Dynamic Lotteries for Multi-group Fairness-Aware Recommendation}.
\newblock \bibinfo{howpublished}{Presented at the 2020 FAccTRec Workshop on Responsible Recommendation}.
\newblock
\showeprint[arxiv]{2009.02590}~[cs.IR]


\bibitem[Sonboli et~al\mbox{.}(2020b)]%
        {sonboli2020opportunistic}
\bibfield{author}{\bibinfo{person}{Nasim Sonboli}, \bibinfo{person}{Farzad Eskandanian}, \bibinfo{person}{Robin Burke}, \bibinfo{person}{Weiwen Liu}, {and} \bibinfo{person}{Bamshad Mobasher}.} \bibinfo{year}{2020}\natexlab{b}.
\newblock \showarticletitle{Opportunistic Multi-Aspect Fairness through Personalized Re-Ranking}. In \bibinfo{booktitle}{\emph{Proceedings of the 28th ACM Conference on User Modeling, Adaptation and Personalization}} (Genoa, Italy) \emph{(\bibinfo{series}{UMAP '20})}. \bibinfo{publisher}{Association for Computing Machinery}, \bibinfo{address}{New York, NY, USA}, \bibinfo{pages}{239–247}.
\newblock
\showISBNx{9781450368612}
\urldef\tempurl%
\url{https://doi.org/10.1145/3340631.3394846}
\showDOI{\tempurl}


\bibitem[S{\"u}hr et~al\mbox{.}(2019)]%
        {suhr2019two}
\bibfield{author}{\bibinfo{person}{Tom S{\"u}hr}, \bibinfo{person}{Asia~J Biega}, \bibinfo{person}{Meike Zehlike}, \bibinfo{person}{Krishna~P Gummadi}, {and} \bibinfo{person}{Abhijnan Chakraborty}.} \bibinfo{year}{2019}\natexlab{}.
\newblock \showarticletitle{Two-sided fairness for repeated matchings in two-sided markets: A case study of a ride-hailing platform}. In \bibinfo{booktitle}{\emph{Proceedings of the 25th ACM SIGKDD International Conference on Knowledge Discovery \& Data Mining}}. \bibinfo{pages}{3082--3092}.
\newblock


\bibitem[Wu et~al\mbox{.}(2022)]%
        {wu2022multi}
\bibfield{author}{\bibinfo{person}{Haolun Wu}, \bibinfo{person}{Chen Ma}, \bibinfo{person}{Bhaskar Mitra}, \bibinfo{person}{Fernando Diaz}, {and} \bibinfo{person}{Xue Liu}.} \bibinfo{year}{2022}\natexlab{}.
\newblock \showarticletitle{A multi-objective optimization framework for multi-stakeholder fairness-aware recommendation}.
\newblock \bibinfo{journal}{\emph{ACM Transactions on Information Systems}} \bibinfo{volume}{41}, \bibinfo{number}{2} (\bibinfo{year}{2022}), \bibinfo{pages}{1--29}.
\newblock


\bibitem[Zehlike et~al\mbox{.}(2022)]%
        {zehlike2022fair}
\bibfield{author}{\bibinfo{person}{Meike Zehlike}, \bibinfo{person}{Tom S{\"u}hr}, \bibinfo{person}{Ricardo Baeza-Yates}, \bibinfo{person}{Francesco Bonchi}, \bibinfo{person}{Carlos Castillo}, {and} \bibinfo{person}{Sara Hajian}.} \bibinfo{year}{2022}\natexlab{}.
\newblock \showarticletitle{Fair Top-k Ranking with multiple protected groups}.
\newblock \bibinfo{journal}{\emph{Information Processing \& Management}} \bibinfo{volume}{59}, \bibinfo{number}{1} (\bibinfo{year}{2022}), \bibinfo{pages}{102707}.
\newblock


\bibitem[Zemel et~al\mbox{.}(2013)]%
        {zemel2013learning}
\bibfield{author}{\bibinfo{person}{Rich Zemel}, \bibinfo{person}{Yu Wu}, \bibinfo{person}{Kevin Swersky}, \bibinfo{person}{Toni Pitassi}, {and} \bibinfo{person}{Cynthia Dwork}.} \bibinfo{year}{2013}\natexlab{}.
\newblock \showarticletitle{Learning fair representations}. In \bibinfo{booktitle}{\emph{Proceedings of the 30th International Conference on Machine Learning (ICML-13)}}. \bibinfo{pages}{325--333}.
\newblock


\bibitem[Zhang and Wu(2017)]%
        {zhang2017anti}
\bibfield{author}{\bibinfo{person}{Lu Zhang} {and} \bibinfo{person}{Xintao Wu}.} \bibinfo{year}{2017}\natexlab{}.
\newblock \showarticletitle{Anti-discrimination learning: a causal modeling-based framework}.
\newblock \bibinfo{journal}{\emph{International Journal of Data Science and Analytics}}  \bibinfo{volume}{4} (\bibinfo{year}{2017}), \bibinfo{pages}{1--16}.
\newblock


\bibitem[Zhu et~al\mbox{.}(2018)]%
        {zhu2018fairness}
\bibfield{author}{\bibinfo{person}{Ziwei Zhu}, \bibinfo{person}{Xia Hu}, {and} \bibinfo{person}{James Caverlee}.} \bibinfo{year}{2018}\natexlab{}.
\newblock \showarticletitle{Fairness-aware tensor-based recommendation}. In \bibinfo{booktitle}{\emph{Proceedings of the 27th ACM International Conference on Information and Knowledge Management}}. \bibinfo{pages}{1153--1162}.
\newblock


\bibitem[Zwicker(2016)]%
        {Zwicker:Voting}
\bibfield{author}{\bibinfo{person}{William~S. Zwicker}.} \bibinfo{year}{2016}\natexlab{}.
\newblock \showarticletitle{Introduction to the Theory of Voting}.
\newblock In \bibinfo{booktitle}{\emph{Handbook of Computational Social Choice}}, \bibfield{editor}{\bibinfo{person}{Felix Brandt}, \bibinfo{person}{Vincent Conitzer}, \bibinfo{person}{Ulle Endriss}, \bibinfo{person}{J{\'{e}}r{\^{o}}me Lang}, {and} \bibinfo{person}{Ariel~D. Procaccia}} (Eds.). \bibinfo{publisher}{Cambridge University Press}, \bibinfo{pages}{23--56}.
\newblock
\urldef\tempurl%
\url{https://doi.org/10.1017/CBO9781107446984.003}
\showDOI{\tempurl}


\end{thebibliography}

\end{document}